\documentclass{article}
\usepackage{graphicx}

\begin{document}
%\draft

%<<<<<<<<<<<<< TITLE >>>>>>>>>>>>>>>%
\title{\bf Kaluza-Klein bubble like structure and celestial sphere in inflationary universe}

%<<<<<<<<<<<<< AUTHOR >>>>>>>>>>>>>>>%
\author{Tetsuya Shiromizu\thanks{Department of Physics, Tokyo Institute of Technology, Tokyo 2-12-1, Japan/E-mail:shiromizu@phys.tiech.ac.jp}, Shinya Tomizawa \thanks{Department of Physics, Osaka City University, 3-3-138, Sugimoto, Sumiyoshiku, Osaka City, Osaka, 113-0033, Japan /E-mail:tomizawa@th.phys.titech.ac.jp : Tel/+81-3-6690-2650},\and Yuki Uchida\thanks{Department of Physics, Tokyo Institute of Technology, 
Tokyo 2-12-1, Oookayama, Meguroku, Tokyo, 152-8551, Japan/E-mail:uchida@th.phys.tiech.ac.jp} and Shinji Mukohyama\thanks{Department of Physics, The University of Tokyo,  Tokyo
113-0033, Japan/E-mail:mukoyama@phys.s.u-tokyo.ac.jp}}

%<<<<<<<<<<<<< DATE >>>>>>>>>>>>>>>%
\date{\today}

%======================================%
%<<<<<<<<<<<<< ABSTRACT >>>>>>>>>>>>>>>%
%======================================%

%<<<<<<<<<<<<< ADDRESS >>>>>>>>>>>>>>>%
%\footnote{Department of Physics, Tokyo Institute of Technology, 2-12-1, Oookayama, Meguroku, Tokyo 113-0033, Japan  E-mail/tomizawa@th.phys.titech.ac.jp : Tel/+81-3-5734-2364 : Fax/+81-3-5734-2745}%

%$^{(1)}$Department of Physics, Tokyo Institute of Technology, 
%Tokyo 2-12-1, Oookayama, Meguroku, Tokyo, 152-8551, Japan\\
%E-mail/tomizawa@th.phys.titech.ac.jp : Tel/+81-3-5734-2364 : Fax/+81-3-5734-2745\\

%$^{(2)}$Department of Physics, The University of Tokyo,  Tokyo
%113-0033, Japan

%$^{(3)}$Jefferson Laboratory of Physics, Harvard University, Cambridge, Massachusetts 02138, USA 

%$^{(4)}$Advanced Research Institute for Science and Engineering, 
%Waseda University, Tokyo 169-8555, Japan%

\maketitle
%\vskip1cm
\begin{abstract}
 We consider five dimensional deSitter spacetimes with a deficit angle
 due to the presence of a closed 2-brane and identify one dimension as
 an extra dimension. From the four dimensional viewpoint we can see that 
 the spacetime has a structure similar to a Kaluza-Klein bubble of
 nothing, that is, four dimensional spacetime ends at the 2-brane. Since
 a spatial section of the full deSitter spacetime has the topology of a
 sphere, the boundary surface surrounds the remaining four dimensional
 spacetime, and can be considered as the celestial sphere. After the
 spacetime is created from nothing via an instanton which we describe,
 some four dimensional observers in it see the celestial sphere falling
 down, and will be in contact with a $2$-brane 
 attached on it. 
\end{abstract}


\clearpage
It is well known that the Kaluza-Klein vacuum is unstable against 
creation of bubbles of nothing in general~\cite{Witten}. The first 
solution found by Witten describes the Kaluza-Klein bubble in 
asymptotically flat spacetimes with a compactified space 
dimension. Recently, extension to asymptotically adS spacetimes was 
considered from the aspect of the adS/CFT correspondence or dynamical 
spacetimes \cite{adS}. The regular, dynamical and non-trivial geometries 
might give us good examination for superstring theory. (See also
Ref. \cite{Ochiai} for similar solution in the context of the
Randall-Sundrum brane world.) On the other hand, a Kaluza-Klein bubble
solution in de Sitter spacetime has not been found.

Consistent de Sitter vacua in superstring theory were just
recently found~\cite{KKLT} and, thus, we now have all kinds of maximally
symmetric spacetimes in superstring theory: Minkowski, de Sitter and
anti-de Sitter spacetimes. Among these three types of spcetimes, a
Kaluza-Klein bubble solution has not been found in and only in de Sitter
spacetime. Therefore, it seems natural to seek a Kaluza-Klein bubble
solution in de Sitter spacetime.

This paper can be considered as a first step towards this
direction. While we shall not find a Kaluza-Klein bubble solution itself
in de Sitter spacetime, we examine properties of a spacetime with 
non-trivial topology, which has structures quite similar to those of a
Kaluza-Klein bubble. 
%Aside from this point, the spacetime structure we
%shall illustrate is by itself interesting. 
First we shall discuss the five 
dimensional de Sitter spacetime with a deficit angle due to the presence
of a closed 2-brane and then find a spacetime structure similar to a
Kaluza-Klein bubble. That is, the four dimensional spacetime ends at the
2-brane if we identify one dimension with an extra dimension. The
construction of full spacetime here is a straightforward extension of
Ref. \cite{Vilenkin} in four dimension to five dimensions. Non-trivial
issue presented here is analysis from the four dimensional point of
view.

We begin with the five dimensional deSitter spacetime in the static
chart 
%===========<Equation>===========%
%
\begin{eqnarray}
ds^2  =  -(1-H^2r^2)dt^2+\frac{1}{1-H^2r^2}dr^2 +r^2(d\theta^2+{\rm sin}^2\theta d\Omega^2_2).
\end{eqnarray}
%
%================================%
Through the double Wick rotation 
%===========<Equation>===========%
%
\begin{eqnarray}
t \to i\chi/H ~~{\rm and}~~\theta \to i\tau +\frac{\pi}{2},
\end{eqnarray}
%
%================================%
we obtain a new metric 
%===========<Equation>===========%
%
\begin{eqnarray}
{}^{(5)}g  =  (H^{-2}-r^2)d\chi^2+\frac{1}{1-H^2r^2}dr^2 +r^2(-d \tau^2+{\rm cosh}^2\tau d\Omega^2_2),
    \label{eqn:5dmetric}
\end{eqnarray}
%
%================================%
where $ 0 \leq r \leq H^{-1}$. If we require the spacetime to be regular 
at $r=H^{-1}$, the coordinate $\chi$ must have the period 
$\chi_p = 2\pi$. As seen later, the spacetime is exactly five dimensional 
deSitter spacetime. On the other hand, if we add a $2$-brane at 
$r=H^{-1}$ then this condition is relaxed and the period can be any 
value smaller than this value: 
%===========<Equation>===========%
%
\begin{eqnarray}
 0 < \chi_p \leq 2\pi. 
\end{eqnarray}
%
%================================%
The equality holds if and only if the $2$-brane tension vanishes or 
there is no $2$-brane. In the following we consider a $2$-brane with 
non-zero tension at $r=H^{-1}$ to make the period $\chi_p$ small enough 
and may regard $\chi$ as a coordinate of a compact dimension. 
Again, as seen later, the spacetime is five dimensional deSitter 
spacetime with a deficit angle.

\begin{figure}[htbp]
\begin{center}
\includegraphics[width=0.4\linewidth]{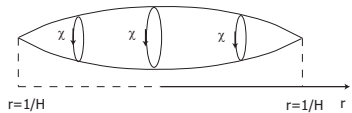}
\end{center}
 \caption{The $2$-dimensional subspace spanned by ($\chi$, $r$). A 
 $2$-sphere is attached on each point and $2$-spheres at ($\chi$, $\pm 
 r$) are identified. 
 \label{fig:pancake}} 
\end{figure}

For $\chi_p<2\pi$, the $2$-dimensional subspace spanned by ($\chi$,
$r$) and extended to the negative $r$ region is like an American 
football as shown in Fig.~\ref{fig:pancake}. (For $\chi_p=2\pi$ it is 
a round sphere.) At each point ($\chi$, $r$) on the surface of the 
football, a $2$-sphere with radius $r\cosh\tau$ is attached and it is 
identified with the $2$-sphere at ($\chi$, $-r$). Hence, if the 
longitudinal angular coordinate of each $2$-sphere is suppressed, a 
$\tau={\rm const.}$ surface has a pancake-like (or convex lens-like)
shape. The thickness of the pancake corresponds to the size of the
compact dimension $\sqrt{H^{-2}-r^2}\chi_p$, and the edge of the
pancake is at $r=H^{-1}$, where a $2$-brane with the topology of
$2$-sphere is attached. While the thickness is independent of the time
$\tau$, the physical radius of the pancake changes as
$H^{-1}\cosh\tau$. 
%===========<Figure 1>===========%
%
%
%================================%

Let us focus on the geometrical structure of timelike hypersurfaces 
$\chi ={\rm const.}$ where we live. The induced metric is 
%===========<Equation>===========%
%
\begin{eqnarray}
{}^{(4)}g= \frac{1}{1-H^2r^2}dr^2 +r^2(-d \tau^2+{\rm cosh}^2\tau d\Omega^2_2).
\end{eqnarray}
%
%================================%
The Riemann tensor of the induced metric becomes 
%===========<Equation>===========%
%
\begin{eqnarray}
{}^{(4)}R_{\mu\nu\alpha\beta}= H^2({}^{(4)} g_{\mu\alpha} {}^{(4)}g_{\nu\beta} -{}^{(4)}g_{\mu\beta}
{}^{(4)}g_{\nu\alpha}).
\end{eqnarray}
%
%================================% 
Since the induced metric has a constant curvature, the spacetime locally 
has maximal symmetry and is locally isometric to a four dimensional 
deSitter (${\rm dS_4}$) spacetime. To investigate the global structure 
we introduce a new coordinate $\sigma$ defined by 
$r=H^{-1}{\rm sin}\sigma$. Then the induced metric is written as 
%===========<Equation>===========%
%
\begin{eqnarray}
H^2{}^{(4)}g=d\sigma^2+{\rm sin}^2\sigma (-d \tau^2+{\rm cosh}^2\tau d\Omega^2_2). 
\end{eqnarray}
%
%================================%
Next we introduce the following coordinate $(T, \theta_1)$ via the 
coordinate transformation 
%===========<Equation>===========%
%
\begin{eqnarray}
{\rm sinh} T & = & {\rm sin} \sigma {\rm sinh} \tau, \nonumber\\
{\rm cosh}T {\rm cos}\theta_1 & = & {\rm cos} \sigma .
\end{eqnarray}
%
%================================%
%================================%
Finally the induced metric becomes deSitter one in the complete chart 
%===========<Equation>===========%
%
\begin{eqnarray}
H^2 {}^{(4)}g=-dT^2+{\rm cosh}^2T (d \theta_1^2 +{\rm sin}^2\theta_1 d\Omega^2_2).
\end{eqnarray}
%
%================================%

To see where $(\tau,\sigma)$ chart covers the deSitter spacetime, it is 
convenient to embed it in a five dimensional Minkowski spacetime as 
%===========<Equation>===========%
%
\begin{eqnarray}
& & Y_0= {\rm sinh}T={\rm sin}\sigma {\rm sinh}\tau ,\nonumber \\
& & Y_\sigma = {\rm cosh}T{\rm cos}\theta_1 = {\rm cos}\sigma, \nonumber \\
& & Y_1 = {\rm cosh}T{\rm sin}\theta_1 {\rm cos}\theta_2 = {\rm sin}\sigma {\rm cosh}\tau 
{\rm cos}\theta_2, \nonumber \\
& & Y_2 = {\rm cosh}T{\rm sin}\theta_1 {\rm sin}\theta_2 {\rm cos}\theta_3 
= {\rm sin} \sigma {\rm cosh}\tau  {\rm sin}\theta_2 {\rm cos}\theta_3, \nonumber \\
& & Y_3 = {\rm cosh}T{\rm sin}\theta_1 {\rm sin}\theta_2 {\rm sin}\theta_3
=  {\rm sin} \sigma {\rm cosh}\tau  {\rm sin}\theta_2 {\rm sin}\theta_3. \nonumber 
\end{eqnarray}
%
%================================%
The deSitter metric is 
%===========<Equation>===========%
%
\begin{eqnarray}
H^2{}^{(4)}g=-dY_0^2+dY_1^2+dY_2^2+dY_3^2+Y^2_\sigma .
\end{eqnarray}
%
%================================%
As shown in Fig. 2, the $\tau={\rm const.}$ slices are cross sections of 
the deSitter hyperboloid $-Y_0^2+Y_1^2+Y_2^2+Y_3^2+Y^2_\sigma =1$ with 
surfaces defined by 
%===========<Equation>===========%
%
\begin{eqnarray}
\Biggl( \frac{Y_0}{{\rm sinh}\tau} \Biggr)^2+ Y_\sigma^2=1. \label{slices}
\end{eqnarray}
%
%================================%
The $\sigma={\rm const.}$ slices are the cross sections of the deSitter 
hyperboloid with $Y_\sigma ={\rm const.}$ surfaces. Hence, the 
coordinate system ($\tau$, $r$) covers the region 
$0\leq Y_{\sigma}\leq 1$. The right boundary $Y_{\sigma}=1$ is just a 
coordinate artifact and the five dimensional spacetime can be 
analytically continued beyond it so that the four dimensional projection 
of the full five dimensional spacetime covers the region $Y_{\sigma}>1$ 
as well, where the deSitter open chart is fitted. On the other hand, we 
can see that the five dimensional spacetime cannot be extended beyond 
the other boundary $Y_{\sigma}=0$. Actually, 
%in the absence of the 
%$2$-brane, 
the five dimensional spacetime is geodesically complete 
%and there is no way to extend it beyond $Y_{\sigma}=0$. 
%With a $2$-brane 
%attached on the edge of the pancake, there is a deficit angle. However, 
%the deficit angle is nothing but a mathematical, macroscopic description 
%of physical, microscopic structures inside the $2$-brane. This means 
%that the five dimensional spacetime cannot be extended beyond 
%$Y_{\sigma}=0$ in this case neither. Thus, there is no four dimensional 
%projection beyond $Y_{\sigma}=0$ in either case. 

Therefore, as depicted in Fig. 2, the regions $Y_{\sigma}<0$ and 
$Y_{\sigma}>0$ represent a similar one with the Kaluza-Klein bubble of 
nothing and the four dimensional world, respectively. The boundary
$Y_{\sigma}=0$ (or $r=H^{-1}$) has the topology of a $2$-sphere and the
area radius $H^{-1}\cosh\tau$, and surrounds both regions. Hence, from  
four dimensional observers' viewpoints, it is interpreted 
that the four dimensional spacetime ends there.

%===========<Figure 1>===========%
%
\begin{figure}[htbp]
\begin{center}
\includegraphics[width=0.3\linewidth]{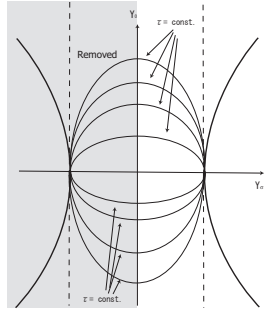}
\end{center}
 \caption{The global structure of the four dimensional induced geometry. 
The figure shows 
the projection onto $Y_0$-$Y_\sigma$ plane. The region of 
$Y{_\sigma}<0$ is removed.} 
\end{figure}
%
%================================%

In order to analyze the global structure we introduce null coordinates 
$u_{\pm}$ defined by 
%===========<Equation>===========%
%
\begin{eqnarray}
u_\pm  =   \tau \pm \int \frac{dr}{r{\sqrt {1-H^2 r^2}}} 
=  \tau \mp \frac{1}{2}{\rm log} \frac{1+{\sqrt {1-H^2r^2}}}{1-{\sqrt {1-H^2r^2}}}. 
\end{eqnarray}
%
%================================%
%Conversely $\tau$ and $r$ are expressed in terms $u_+$ and $u_-$ as 
%===========<Equation>===========%
%
%\begin{eqnarray}
%\tau = \frac{1}{2}(u_+ + u_-)
%\end{eqnarray}
%
%================================%
%and
%===========<Equation>===========%
%
%\begin{eqnarray}
%r= \frac{1}{H}\frac{1}{{\rm cosh} \frac{u_+ - u_-}{2} }.
%\end{eqnarray}
%
%================================%
In this null coordinate system the induced metric becomes 
%===========<Equation>===========%
%
\begin{eqnarray}
{}^{(4)}g=H^{-2} \Biggl[-\frac{1}{{\rm cosh}^2 \frac{u_+ - u_-}{2}} du_+ du_- +R^2(u_+,u_-)d\Omega^2_2   \Biggr]
\end{eqnarray}
%
%================================%
where
%===========<Equation>===========%
%
\begin{eqnarray}
R(u_+,u_-)= \frac{{\rm cosh \frac{u_+ + u_-}{2}}}{{\rm cosh}  \frac{u_+-u_-}{2}}.
\end{eqnarray}
%
%================================%
The expansion rate of out/in-going null geodesic congruence is 
%===========<Equation>===========%
%
\begin{eqnarray}
\theta_\pm := \frac{\partial {\rm ln}R}{\partial u_\pm} 
=\frac{1}{2}\Biggl[\frac{{\rm sinh}\frac{u_+ + u_-}{2}}{{\rm cosh} \frac{u_+ + u_-}{2}} 
\mp \frac{{\rm sinh}\frac{u_+ -u_-}{2}}{{\rm cosh} \frac{u_+ - u_-}{2}}  \Biggr]. 
\end{eqnarray}
%
%================================%
Since the spacetime is locally a four dimensional deSitter spacetime, 
the structure in terms of the expansion rates is simple as seen in 
Fig. 3. Each expansion $\theta_{\pm}$ vanishes on the null surface 
$u_{\mp}=0$, is positive in the region $u_{\mp}>0$ and is 
negative in the region $u_{\mp}<0$. We note that $Hr=1/{\rm cosh}\tau$ 
and $Y_1^2+Y_2^2+Y_3^2=1$ are satisfied on the $\theta_{\pm}=0$ 
surfaces. 
%It is also straightforward to confirm this structure in any 
%other coordinate systems by calculating the Misner-Sharp energy 
%\cite{MS,Hayward}. As easily noticed and as should be, the area radius 
%$H^{-1}R$ of the horizons at $u_{\pm}=0$ is $H^{-1}$.  

From Fig. 3, one can easily see that, from the view point of 
observers sitting at the center $R=0$ (i.e. $x=-\infty$) of the 
spherical symmetry, the boundary $x=0$ is 
always beyond the cosmological horizons $u_{\pm}=0$. 
%This is, of course, 
%consistent with the previously stated fact that the radius of the 
%pancake is $H^{-1}\cosh\tau$. Indeed, the radius of the pancake is 
%larger than the horizon radius $H^{-1}$ except at $\tau=0$, where those 
%two coincide. Moreover, as we shall see later in this paper, the region 
%with negative $\tau$ is continued to an Euclidean solution. Hence, the 
%observer is always outside the causal future of the spacetime boundary $x=0$ 
%in the Lorentzian region. In these senses, observers sitting exactly at 
%the center $R=0$ do not notice the existence of the spacetime boundary $x=0$. 
On the other hand, freely-falling observers off the center can be 
affected by the boundary. For example, let us consider 
comoving observers in the flat-chart covering the region $u_->0$. As 
easily seen from Fig. 3, some observers (the observer 1 in Fig. 3) will 
reach the edge $x=0$ in finite affine time if the deviation from the 
center of spherical symmetry is large enough. For a small but non-zero 
deviation, the observer (the observer 2 in Fig. 3) will not reach the 
edge but will be in the causal future of the bubble in Lorentzian region 
after some time.

So far, we investigated the spacetime structure from the four
dimensional viewpoint. Let us now consider the extra dimension and the
five dimensional structure. The size of the extra dimension is given by
$\sqrt{H^{-2}-r^2}\chi_p$ in the region covered by the coordinate system
adopted in (\ref{eqn:5dmetric}). This expression is equivalent to
$Y_{\sigma}\chi_p$ and, thus, can be extended to the open universe
region $Y_{\sigma}>1$. Hence, the size of the extra dimension is not
bounded from above in the open universe region. It is, thus, natural to
ask what the ratio of the size of extra dimension to the size of
the four dimensional universe is. We shall see below that the ratio is
roughly $\chi_p/4$, which can be made arbitrarily small by fine-tuning
the $2$-brane tension.

\begin{figure}[htbp]
\begin{center}
\includegraphics[width=0.3\linewidth]{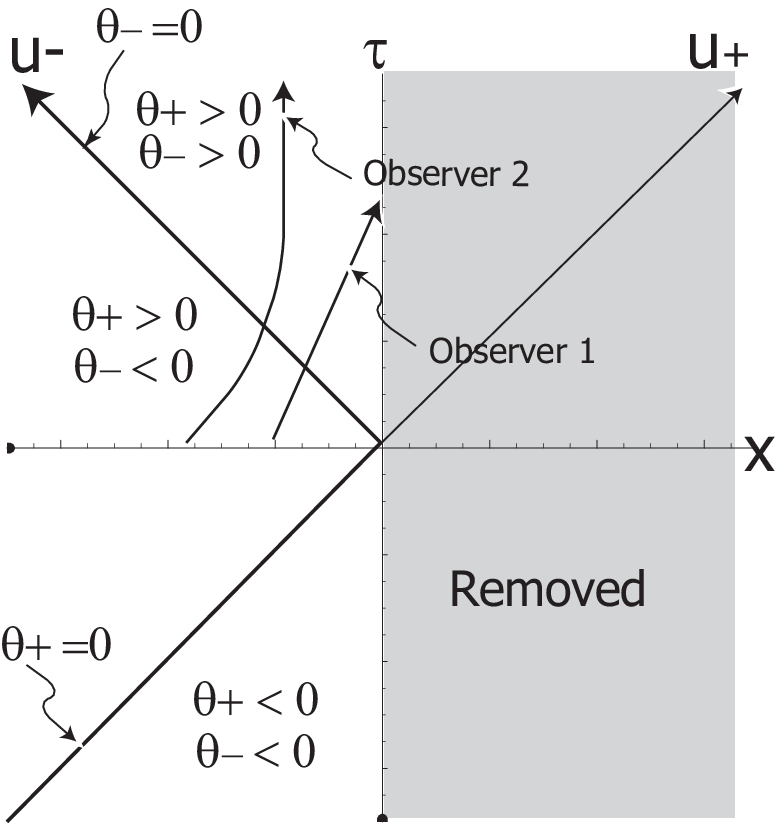}
\end{center}
\caption{
 The signature distribution of $\theta_+$ and $\theta_-$ in $\tau$-$x$ 
 plane. Only the region $x=\frac{1}{2}(u_+-u_-)<0$ is considered since 
 the other region $x>0$ is removed. The surface $\theta_{\pm}=0$ 
 corresponds to $\tau=\pm x<0$ and separates the spacetime into two 
 regions, that is, the region $\theta_{\pm}>0$ (or $\tau>\pm x,x<0)$ and 
 the region $\theta_{\pm}<0$ (or $\tau<\pm x,x<0$). 
 }
\end{figure}

The metric of the full five-dimensional spacetime, which is locally
isometric to the deSitter spacetime, is written as
%===========<Equation>===========%
%
\begin{eqnarray}
 H^2{}^{(5)}g & = & 
 -dT^2+{\rm cosh}^2T (d \theta^2_1+{\rm cos}^2\theta_1 d\chi^2
 + {\rm sin}^2 \theta_1 d\Omega^2_2 ) \nonumber \\
   & = & -dT^2+{\rm cosh}^2T ds^2_{\rm S^4}. 
\end{eqnarray}
%
%================================%
Here, $ds^2_{\rm S^4}$ is the metric of $S^4$ and is transformed to the
standard form by 
%===========<Equation>===========%
%
\begin{eqnarray}
 \cos\tilde{\theta}_1 & = & \cos\theta_1\sin\chi, \nonumber\\
 \sin\tilde{\theta}_1\cos\tilde{\theta}_2 & = & \cos\theta_1\cos\chi, 
  \nonumber\\
 \sin\tilde{\theta}_1\sin\tilde{\theta}_2 & = & \sin\theta_1,
\end{eqnarray}
%
%================================%
so that  
$d\theta^2_1+\cos^2\theta_1d\chi^2+\sin^2\theta_1d\Omega^2_2 
=d\tilde{\theta}_1^2+\sin^2\tilde{\theta}_1 
(d\tilde{\theta}_2^2+\sin^2\tilde{\theta}_2 d\Omega^2_2)$. 
Hence, without the deficit angle due to the $2$-brane, the five
dimensional spacetime would be the maximally extended de Sitter
spacetime not only locally but also globally. On the other hand, we
shall see below that the presence of the $2$-brane and the deficit angle
introduces an intriguing structure. The $2$-brane is located at
$\theta_1 = \pi /2$, where the extra dimension of $S^1$ shrinks.

In order to visualize the five dimensional spacetime structure in the
presence of the closed $2$-brane and the deficit angle
($2\pi-\chi_p>0$), it is convenient to go to a flat chart. 
Note, however, that the same structure looks differently
for different choices of the origin of the flat chart since the presence
of the $2$-brane breaks the global translational invariance. Here, we
choose the origin of the flat chart $(t_1,x_1,x_2,x_3,x_4)$ as 
%===========<Equation>===========%
%
\begin{eqnarray}
 t_1 & = & \ln(\sinh T + \cosh T\cos\tilde{\theta}_1), \nonumber\\
 x_1 & = & \frac{\cosh T\sin\tilde{\theta}_1\cos\tilde{\theta}_2}
  {\sinh T + \cosh T\cos\tilde{\theta}_1}, \nonumber\\
 x_2 & = & \frac{\cosh T\sin\tilde{\theta}_1\sin\tilde{\theta}_2
  \cos\tilde{\theta}_3}{\sinh T + \cosh T\cos\tilde{\theta}_1},
  \nonumber\\
 x_3 & = & \frac{\cosh T\sin\tilde{\theta}_1\sin\tilde{\theta}_2
  \sin\tilde{\theta}_3\cos\tilde{\theta}_4}
  {\sinh T + \cosh T\cos\tilde{\theta}_1}, \nonumber\\
 x_4 & = & \frac{\cosh T\sin\tilde{\theta}_1\sin\tilde{\theta}_2
  \sin\tilde{\theta}_3\sin\tilde{\theta}_4}
  {\sinh T + \cosh T\cos\tilde{\theta}_1}. 
\end{eqnarray}
%
%================================%
In this flat chart, the $2$-brane is located at the intersection of
$x_1=0$ and $x_2^2+x_3^2+x_4^2=1+e^{-2H t_1}$. For a deficit angle
larger than $\pi$ or, equivalently, $\chi_p<\pi$, the spacetime region
is restricted by \footnote{  
In the above argument we have assumed that the deficit angle is larger
than $\pi$. If the deficit angle were smaller than $\pi$ (and thus
$\chi_p>\pi$) then it would be convenient to change the origin of $\chi$
so that $-\chi_p/2\leq\chi+\pi/2\leq\chi_p/2$. This would correspond to
the change of the origin of the flat chart to the opposite point on the
$S^4$. With this change, the inequality in (19) would be
opposite and $\chi_p$ in the right hand side would be replaced by the
deficit angle  $2\pi-\chi_p$. Therefore, in the case of a large deficit
angle ($\chi_p<\pi$) the spacetime is the outside of a pancake-shaped
region but not the inside. In the limit of zero deficit angle
($\chi_p\to 2\pi$) the maximally extended de Sitter spacetime is
recovered as the thickness of the pancake-shaped region vanishes. }
%===========<Equation>===========%
%
\begin{eqnarray}
|x_1| & \leq &
 \frac{\sqrt{1+(1+e^{-2Ht_1}-x_2^2-x_3^2-x_4^2)\tan^2(\chi_p/2)}-1}
 {\tan(\chi_p/2)} \nonumber\\
 & = & \frac{\chi_p}{4}(1+e^{-2Ht_1}-x_2^2-x_3^2-x_4^2) + O(\chi_p^2),
  \label{eqn:pancake}
\end{eqnarray}
%
%================================%
where we have chosen the origin of $\chi$ so that 
$-\chi_p/2\leq\chi-\pi/2\leq\chi_p/2$. Therefore, the
$5$-dimensional spacetime has a pancake-like (or convex lens-like) shape
as already discussed. The expression (19) also makes it 
evident that the ratio of the size of the extra dimension (the thickness
of the pancake) to the size of the four dimensional universe (the radius
of the pancake) is roughly $\chi_p/4$.

%
%As already stated, the same structure looks different for different
%choices of the origin of the flat chart. For example, we could choose
%the opposite point on the $S^4$. This would correspond to changing the
%origin of $\chi$ so that $-\chi_p/2\leq\chi+\pi/2\leq\chi_p/2$. In this
%flat chart, the 

\begin{figure}[htbp]
\begin{center}
\includegraphics[width=0.3\linewidth]{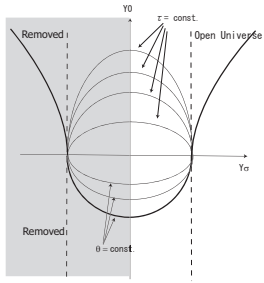}
\end{center}
\caption{Creation of the universe.}
\end{figure}

%Unlike Witten's original Kaluza-Klein bubble solution without a 
%cosmological constant, the current solution 
%we have discussed does not necessarily mean semi-classical 
%instabilities. Actually, we have not found a Euclidean bounce solution 
%representing a quantum-mechanical decay of a classically stable, regular 
%spacetime to the spacetime with the current solution we have
%discussed. 

%Instead of a bounce solution, we can easily find a no-boundary instanton
%by analytically continuing the spacetime to a deSitter spacetime with a 
%deficit angle. Thus, in five dimensional theories with a positive cosmological  
%constant, the spacetime can be created from
%nothing via the instanton. 

The spacetime discussed here can be created from nothing via an instanton. 
The instanton is obtained simply by analytic continuation on the 
$\tau=0$ surface. In the Euclidean regime 
($\tau \to -i (\tau_E-\pi/2), Y_0 \to i Y_E , T \to i(T_E-\pi/2)$), the 
five dimensional Euclidean geometry is locally isometric to the 
$5$-sphere $S^5$ as 
%===========<Equation>===========%
%
\begin{eqnarray}
H^2{}^{(5)}g_E=dT^2_E+{\rm sin}^2T_E ds^2_{S^4}=ds^2_{S^5},
\end{eqnarray}
%
%================================%
while Eq. (\ref{slices}) becomes 
%===========<Equation>===========%
%
\begin{eqnarray}
\Biggl( \frac{Y_E}{{\rm cos}\tau_E} \Biggr)^2+ Y_\sigma^2=1.
 \label{slices-Euclid}
\end{eqnarray}
%
%================================%
Hence $\theta={\rm const}$ surfaces are cross sections of the four 
dimensional unit-sphere $Y_E^2+Y_1^2+Y_2^2+Y_3^2+Y_{\sigma}^2=1$ with 
the surface defined by Eq. (\ref{slices-Euclid}) (See Fig. 4). The creation 
of the spacetime with 2-brane at $\tau=0$ because 
$\tau=0$ surface is momentary static. Note that the path depicted in 
Fig. 4 corresponds to a quantum creation of the whole universe with a
closed 2-brane. As we have already mentioned, the region 
$Y_\sigma \geq 1$ can be regarded as an open universe. So the current 
path may be interpreted as creation of the open inflationary universe 
with a 2-brane. (See Ref. [6] for creation of open 
universes via a singular instanton. In Ref. [7,8] a 
regular instanton similar to our instanton was proposed.)

Let us summarize our study. In a five dimensional theory admitting a
positive cosmological constant, we considered a spacetime with a closed
2-brane. Despite the fact that it is locally a de Sitter spacetime, a
large enough deficit angle due to the presence of the $2$-brane
introduces an intriguing structure. Depending on observers, the
spacetime can be effectively four dimensional at low energy. Indeed,
while the size of the extra-dimension is not bounded from above in
general, its ratio to the size of the four dimensional universe can be
made as small as one likes by fine-tuning the $2$-brane tension. If the
deficit angle exceeds $\pi$ then the spacetime has a pancake-like (or
convex lens-like) shape with the upper and lower surfaces
identified. All observers are confined inside the pancake-shaped region
surrounded by the closed $2$-brane.

The edge of the pancake mimics the surface of the Kaluza-Klein
bubble of nothing and all observers in the pancake are outside the
``bubble''. Hence, it is the boundary of the four dimensional
spacetime.

\section*{Acknowledgements}

The work of TS was supported by Grant-in-Aid for Scientific 
Research from Ministry of Education, Science, Sports and Culture of 
Japan(No.13135208, No.14740155 and No.14102004). 
The works of ST and YU were supported by the 21st Century COE Program at 
TokyoTech "Nanometer-Scale Quantum Physics" supported by the Ministry of 
Education, Culture, Sports, Science and Technology. SM is supported by 
NSF grant PHY-0201124.

%\vskip 1cm 

\end{document}